\documentstyle[12pt,epsf]{article}
\def\lsim{\:\raisebox{-0.5ex}{$\stackrel{\textstyle<}{\sim}$}\:}

\def\goes{\rightarrow}
 
         \def\mz{M_{Z}}

                 \def\tb{\tan \beta}
                 \def\mg{M_{\widetilde g}}
                 \def\msq{m_{\widetilde q}}

                 \def\mser{m_{{\widetilde e}_R}}
                 \def\msnu{m_{\widetilde \nu}}

         \def\su{\widetilde u}
         \def\sd{\widetilde d}
         
         \def\sql{{\widetilde q}_L}
         \def\sqr{{\widetilde q}_R}

         \def\slel{{\widetilde e}_L}
         \def\sler{{\widetilde e}_R}

         \def\snu{\widetilde \nu}

         \def\l2p{\lambda^{\prime\prime}}
\def\prd{{\it Phys. Rev.} {\bf D}}
\def\prl{{\it Phys. Rev. Lett.}}
\def\plb{{\it Phys. Lett.} {\bf B}}
\def\npb{{\it Nucl. Phys.} {\bf B}}
\def\zpc{{\it Z. Phys.} {\bf C}}
\def\ijmpa{{\it Int. J. Mod. Phys.} {\bf A}}
\def\mpla{{\it Mod. Phys. Lett.} {\bf A}}

\def\pr{{\it Phys. Reports}}
         \def\etal{{\it et al}}
\begin{document}
                 \setcounter{page}{0}
                 \thispagestyle{empty}
\begin{flushright} 
BU-TH/96-2 ~;~ IISc-CTS-7/96 ~;~ TIFR/TH/96-21 \\ {\bf hep-ph/9605460} \\ 
\end{flushright}
\begin{center} 
{\large\bf 
             CAN THE ALEPH FOUR-JET EXCESS BE EXPLAINED IN A 
             SUPERSYMMETRIC MODEL WITH R-PARITY VIOLATION?} \\
\vskip 0.2in
                       Dilip Kumar Ghosh$^a$ \\
          {\it Department of Physics, University of Bombay, \\
          Vidyanagari, Santa Cruz (East), Mumbai 400 098, India.} \\
\bigskip
                       Rohini M. Godbole
\footnote{on leave of absence from the Department of Physics, University
of Bombay, Mumbai, India.}$^{,b}$  \\
      {\it Centre for Theoretical Studies, Indian Institute of Science, \\
                      Bangalore 560 012, India.} \\
\bigskip
                       Sreerup Raychaudhuri$^c$ \\ 
{\it 
Theoretical Physics Group, Tata Institute of Fundamental Research, \\ 
          Homi Bhabha Road, Mumbai 400 005, India.} \\
\bigskip\bigskip
                       {\large\bf Abstract}
\end{center}
We investigate the possibility that the excess of four-jet events in
$e^+e^-$ collisions at LEP-1.5 which has been reported by the ALEPH
Collaboration could be due to the production of charginos or
neutralinos, followed by their decay into quark jets through baryon
number-violating ($\lambda^{\prime\prime}$) couplings. An 
estimate at the parton level shows, however, 
that these events cannot be due to
neutralinos because of the low cross-section, and is unlikely 
to be due to the production of chargino pairs because of the largely
different event shapes. \\

\noindent
Pacs. Nos: 14.80.Ly, 13.65.+i, 12.60.Jv \\

\bigskip

{\footnotesize \noindent Electronic address: 
    ~$^a$dghosh@theory.tifr.res.in; 
~ ~ ~$^b$rohini@cts.iisc.ernet.in; 
~ ~ ~$^c$sreerup@theory.tifr.res.in }
\newpage
\begin{center}
{\bf 1. INTRODUCTION.}
\end{center}

In the last few months, considerable interest has been generated by
fresh results coming from the LEP Collider at CERN. Upgradation of
the collision energy to 130/136 GeV has given rise to expectations
that one might see new physics effects at this energy. However, most
of the results obtained till now at this LEP-1.5 collider are
consistent \cite{LEP-1.5_results} with the predictions of the
Standard Model (SM) and merely lead to new constraints on physics
beyond it. There is, however, one notable exception to this trend.
The ALEPH Collaboration has reported \cite{ALEPH_4jet} an excess over
the SM prediction in the $e^+e^- \goes$ {\it four-jets} channel. The
number of events in the data sample of 5.7 pb$^{-1}$ which
satisfy the rather stringent criteria imposed for this search is 16,
which is significantly in excess of the 8.6 events predicted by the
SM. Of these 16 events, 9 have a four-jet invariant mass of about 105
GeV. The rest are consistent with the SM background.

This observation is rather hard to interpret. Unless the observed result 
is due to a statistical fluctuation, which is somewhat remote in
view of the low estimated probability of 10$^{-4}$, it seems natural
to assume that both the measurement and the SM prediction are correct 
and hence, the four-jet excess is a genuine new physics effect. While 
this may seem to be unduly
optimistic at this preliminary stage of the upgraded LEP runs, it is
nevertheless amusing to probe new physics scenarios which could lead
to a four-jet excess. One such scenario is discussed in this article,
namely, supersymmetry with violation of $R$-parity as a possible
candidate for this effect. 

The basic idea expanded in our work is rather simple. We consider two
possibilities within the framework of the Minimal Supersymmetric
Standard Model (MSSM) with $R$-parity violation
\cite{Rviolation_reviews,Rviolation_bounds}.
\begin{enumerate} 
\item The lightest {\it neutralino}, which is assumed to be the 
lightest
\footnote{This is natural in a $R$-parity conserving scenario, but
is not essential if $R$-parity is violated; however, we keep
this configuration since it involves the minimum change from the
familiar MSSM scenarios.}
supersymmetric particle (LSP), may have mass around 50 --- 55 GeV and
be pair-produced in $e^+e^-$ collisions at LEP-1.5. If $R$-parity is
not conserved, these neutralinos could decay \cite{LSP_decays} and
the invariant mass of the decay products of each would peak around
50 --- 55 GeV. The sum of these invariant masses would then peak
around 105 GeV.  If $R$-parity is violated through baryon
number-violating couplings ($\l2p$), the neutralinos will decay into
three quark jets each, which could then merge to give the distinctive
four-jet signals.
\item The lighter {\it chargino} (which is assumed to be 
heavier than the lightest
neutralino) may have mass around 50 --- 55 GeV and be pair-produced
similarly at LEP-1.5. As before, when the charginos decay, the sum of
invariant masses of the decay products of both would be expected to
peak in or around  105 GeV. We then envisage the decay of {\it each}
chargino to a neutralino (LSP) and a pair of quark jets; this LSP
then decays through $\l2p$ couplings into three jets.  Thus one has
ten jets in all, which could then merge to give four-jet signals. We
could also, in principle, expect signals in channels with more jets,
though, as it turns out, these are not really significant.
\end{enumerate} 
Recently, it has been pointed out \cite{Dreiner} that it is possible
for the chargino to decay {\it directly} into jets through $R$-parity
violating $\l2p$ couplings without the intermediacy of a neutralino
as considered in 2. This is an exciting possibility but we do not 
pursue it here; our assumption being that the neutralino is suficiently
lighter than the chargino and that the relevant 
$\l2p$ coupling is too small for the direct chargino decay to compete 
with the usual MSSM decay mode. This is consistent with our philosophy 
of keeping the changes from the conventional MSSM scenario to a minimum.

We analyse the above signals using a parton-level Monte Carlo event
generator to scan the MSSM parameter space. The use of a parton-level
generator has two advantages: ($a$) it is quick, so that a detailed
study of the parameter space is possible; and ($b$) the analysis is
relatively simple so that one can focus on the basic physics issues.
On the other hand, this approach has the obvious drawback that the
algorithms used to analyse jets are necessarily crude. Hence our
results should be considered diagnostic only. The issues which have
been addressed in this work are confined to checking if
the processes under consideration are at all viable and (if
such is the case) to identify the relevant part of the MSSM
parameter space.  In case of a viable signal, our investigation
could, in principle, aid further studies of the process in which the 
analysis of jets is done in greater detail.

Before we pass on to the details of our analysis, it would be
appropriate to discuss in greater detail the nature of the multijet
events we are analysing. The events seen by the ALEPH Collaboration
\cite{ALEPH_4jet} consist of spherically distributed multijet events
where the event shapes are consistent with a purely hadronic final
state.  All events satisfy the requirement that the net visible
energy of the jets is at least 70 \% of the centre-of-mass energy at
LEP-1.5.  Moreover there are practically no displaced vertices in the
microvertex detector, indicating that the jets originate from light
quarks or gluons rather than $b$-quarks.
 
To select these events out of the SM multijet background, the ALEPH
Collaboration has imposed a number of kinematic and other cuts 
on the jets. The
principal background comes from the process $e^+e^- \goes q \bar{q}$
followed by gluon radiation from the quarks. Most of the gluon jets,
however, will be soft and much of this background is eliminated when
the following selection criteria are imposed:
\begin{enumerate}
\item At least eight charged particle tracks must be reconstructed with
at least four hits in the time projection chamber, with a polar angle
$\theta$ with respect to the beam such that $\mid \cos \theta \mid < 0.95$
and originating within a cylinder of length 20 cm and radius 2 cm coaxial
with the beam and centred at the nominal collision point.
\item The scalar sum of the charged particle momenta must exceed 10\%
of the centre-of-mass energy.
\item Radiative $Z^0$ returns are removed by requiring the missing
momentum measured along the beam direction to be smaller than 
$0.75 (m_{vis} - 90~{\rm GeV})$, where $m_{vis}$ is the invariant mass of 
the system formed by all energy-flow particles.
\item Events with fewer than four jets are rejected.  
\item None of the jets (in the four jets sample) contains more than
80\% electromagnetic energy.
\item The dijet invariant mass of each pair of jets in the {\it
four-jet} sample (6 pairs in all) is not less than 25 GeV;
\item The sum of pairs of jet invariant masses is not less than 10 GeV;
\item Each pair of jets has a minimum of 10 charged tracks between them.
\end{enumerate}
It should be noted that the ALEPH Collaboration observe no six-jet
events and all five-jet events have been converted into four-jet
events by merging the pair of jets which has the minimum dijet
invariant mass. They report the observation of 16 events which
satisfy all these criteria against a SM prediction of 8.6 events.
What is even more interesting, however, is the distribution in four-jet
invariant mass. If the jets are numbered 1,2,3,4 and
$\Delta M \equiv min \mid M_{ij} - M_{kl} \mid$ out of the combinations
$ij;kl =$ 12;34 or 13;24 or 14;23 respectively, then for 9 of these 
events the four-jet invariant mass, defined as the sum $\Sigma M \equiv 
(M_{ij} + M_{kl})$ for the combination $ij;kl$ which yields $\Delta M$, 
lies within 102.1 --- 108.4 GeV.  The predicted SM background in this bin
is about 0.8 events for 5.7 pb$^{-1}$ luminosity. The probability of
this accumulation being due to a fluctuation in the SM background is 
as estimated to be as low as $1.1 \times 10^{-4}$ \cite{ALEPH_4jet}.

In section 2 we discuss the decay of the LSP in the $R$-parity
violating model of our choice and explain our techniques for
analysing multijet signals. Section 3 analyses the possibility that
the four-jet anomaly may be due to neutralinos. In section 4 we make
a similar discussion for charginos. Finally, section 6 contains a
summary of our results and conclusions.

\begin{center}
{\bf 2. LSP DECAY WITH BARYON NUMBER VIOLATION.}
\end{center}

The crucial feature of our analysis is the decay of the neutralino
(assumed to be the LSP) into jets through $R$-parity violating couplings. 
These
couplings can, in general, be of two kinds: those that violate lepton
number and those that violate baryon number. If {\it both} are
present, the theory predicts a large width for proton decay which is
inconsistent with current data \cite{proton_decay}. Accordingly, one
has to assume that either lepton or baryon number-violating couplings
are present, but not both. For the purposes of this work, we assume
that there are no lepton number-violating couplings and concentrate
on baryon number-violation only. The baryon number-violating term in
the superpotential has the form
\cite{Rviolation_reviews}
\begin{equation}
{\cal W}_{B \!\!\!\!/} = 
\sum_{ijk} \l2p_{ijk} {\hat U}^c_i {\hat D}^c_j {\hat D}_k^c
\end{equation}
where ${\hat U}^c, {\hat D}^c$ are chiral superfields containing the
right-handed $u,d$-quarks and the indices $i,j,k$ run over the three
quark generations. This leads to the interaction Lagrangian
\begin{equation}
{\cal L}^{int}_{B \!\!\!\!/} = - \sum_{ijk} \l2p_{ijk}
\Big[ \su^*_{Ri} {\bar d}_{Rk} d_{Lj}^c
    + \sd^*_{Rj} {\bar u}_{Ri} d_{Lk}^c 
    + \sd^*_{Rk} {\bar d}_{Rj} u_{Li}^c  \Big] + H.c.
\end{equation}
Though colour indices are not explicitly shown, the interaction term
must be a colour singlet; this requires the $\l2p_{ijk}$ to be
antisymmetric in the last two indices.

One can now envisage the decay of a LSP into a quark and an off-shell
squark which then goes to a pair of quarks through the above baryon
number-violating couplings \cite{LSP_decays}. 
There are three possible diagrams
corresponding to the three terms in the interaction Lagrangian and
these are shown in Figure 1. For this study, we assume that the
coupling $\l2p_{212}$ is dominant and that the others may be
neglected. This is not an essential requirement of the theory, but is
the simplest option \cite{BaerKaoTata}. However, it is important to
note that the decay width is proportional to the square of this
coupling $\l2p_{212}$ only, so that the coupling {\it cancels} out of
numerator and denominator in the branching ratio. In any case, since
this is the only decay mode of the LSP in the scenario under
consideration, the branching ratio of the LSP to jets is unity. It is
also important to note that the $\l2p_{212}$ coupling ensures that
there are no $b$-jets in the final sample, which is what the ALEPH
Collaboration finds\footnote{Actually {\it one} of the 16 events does
contain \cite{ALEPH_4jet} $b$-jets, but this could well arise from
the QCD background.}.

Till the present date, no direct {\it search} has been made at LEP-1
for $R$-parity violating signals in the presence of $\l2p$ couplings,
though corresponding searches have been made assuming the presence of
$\lambda$ couplings \cite{ALEPH_rviol} and strategies have been
discussed for $\lambda^\prime$ and $\l2p$ couplings
\cite{barger_rviol}. Thus, there are no direct bounds from LEP-1 data
on MSSM parameters in this scenario. Such a study is, however,
possible, and is in progress \cite{L3andUs}.  Moreover, since the LSP
decays (into jets), the usual missing energy signals are not expected
to be seen. In view of this, the only constraint coming from LEP-1
data is the requirement that the contribution of LSP pair-production
to the total $Z$-width should be consistent with the SM prediction
and the experimental error \cite{LEP-1_bounds}. There is also a 
strong bound from the nonobservation of charginos at LEP-1 leading to
the requirement that the mass of the chargino should exceed 45 GeV;
however, this constraint affect roughly the same region of 
parameter space as the previous one, since the chargino contribution
to the $Z$-total width is large.  Furthermore, for $\mg
\lsim 150$ GeV, one should have \cite{BaerDreesTata} seen spectacular
multijet signals at LEP-1, which is not the case.  However, one
certainly needs to consider a somewhat larger allowed region in the
MSSM parameter space than is the case for $R$-parity conserving
models.

Since the decay of the LSP will lead to final states with several
quark jets, we now explain our method of determining the number of
distinct jets. As explained before, we use a simple parton-level
Monte Carlo event generator which is unable to simulate the details
of the hadronic fragmentation. We therefore make the somewhat crude
approximation that the direction of the parent parton is the same as
that of the thrust axis of the resulting jet and that the hadronic
material is confined to a cone around it. Thus, if the directions of
two partons which engender jets are separated by $\sqrt{\Delta \eta^2
+ \Delta \phi^2} \equiv \Delta R \leq 0.7$, we consider the jets to
have merged. While we recognise that this approximation is a somewhat
crude one, it is a commonly used rule of thumb in analyses of jets
produced around the electroweak scale, at least at hadron colliders
\cite{cone_algorithm}. We have also considered the Durham algorithm 
(used by the ALEPH Collaboration) for
the jet-merging procedure, merging partons with $y < 0.008$
rather than the fixed cone algorithm discussed above, but the final
results do not change by more than a few percent from the results
presented in this work (using the cone algorithm).
Once a pair of jets is considered to have
merged, the momentum and thrust axis of the resultant jet are
constructed simply by vectorially adding the three-momenta of the
original jets. 

In our analysis we consider the production of a LSP pair or a
chargino pair and their subsequent decay to multijet final states.
These are allowed to pass through the `jet'-merging algorithm described
above and the reconstructed `jets' (by which we mean one or more 
partons, or more specifically, quarks) are passed successively through the
following kinematic cuts, which more-or-less follow those used by the 
ALEPH Collaboration in their analysis:

\begin{enumerate}
\item The jets
should have rapidity less than 3, since this roughly corresponds to the 
polar angle cut used by the ALEPH Collaboration. It turns out,
however, that most of the jets which contribute to the final signal
have rapidity $\leq 2$, so we could easily accommodate a more
stringent rapidity cut without affecting the signal. This 
may ultimately be required in view of the lower detection efficiency for
high rapidities.
\item The scalar sum of momenta of the jets in the final analysis must be 
greater than 0.1 of the $e^+e^-$ centre-of-mass energy, {\it i.e.} 13 GeV;
\item The sample should contain four or more jets only. All five-jet
events are converted into four-jet events by merging the pair of jets
with the minimum invariant mass. Events with six or more are jets are 
counted separately.
\item The dijet invariant mass of each pair of jets in the four-jet
sample is greater than 25 GeV.
\end{enumerate}

In the absence of detector simulation, calorimetic cuts cannot be 
imposed. Moreover, in our parton-level event generator, we are unable 
to impose constraints which involve 
the invariant masses of individual jets or the multiplicity of charged 
tracks, since that would be possible only in a simulation which takes 
into account the hadronisation processes. Accordingly, we 
make the {\it ad hoc} 
assumption, in the following discussion, that the supersymmetric signal 
is reduced by 40\% by cuts of this nature. The choice of this factor is
guided by the reduction in the signal and background presented by
the ALEPH Collaboration, which are 27\% and 47\% respectively. While we 
recognise that this is a rather crude approximation, it is not likely to 
affect the conclusions of our study, as we shall see. It should be noted 
that we have not included this reduction factor in the kinematic 
distributions shown in this work;
these would accordingly be reduced and probably 
smeared further by application of these 
cuts in a simulation which takes account of fragmentation. \\

\begin{center}
{\bf 3. FOUR-JET EXCESS FROM NEUTRALINO PAIR-PRODUCTION.}
\end{center}

We first consider the scenario when the four-jet signal is due to the
production of a pair of neutralinos (LSP's). For this, one requires
the mass of the LSP to lie in the range allowed by LEP-1 data and
accessible to LEP-1.5. This automatically restricts our analysis to a
limited region in the MSSM parameter space obtained by variation of
gluino mass ($\mg$), Higgsino mixing parameter ($\mu$) and the ratio
of vacuum expectation values of the two Higgs doublets ($\tan
\beta$).  Dependence on the squark mass is minimised by considering
the gluino mass evaluated at the electroweak scale, {\it i.e.}
$\mg(\mz)$. Of course, in this work, we have made the assumption of 
gaugino
mass unification at a high scale, which enables us to use the gluino
mass as a parameter of the electroweak gaugino sector; however, the 
analysis would be unchanged if we
relaxed this hypothesis and used instead the soft-SUSY breaking parameter
$M_2$ in the $SU(2)$ sector.

The cross-section for production of a pair of neutralinos has been
calculated in the MSSM, in terms of the above parameters, by a number
of authors \cite{neutralino_production}.  We have checked that our
cross-sections are consistent with theirs, both analytically and
numerically.  It may be noted that neutralinos are produced through
$s$-channel $Z$ exchanges as well as $t,u$-channel $\slel,\sler$
exchanges and hence are rather sensitive to the masses of the
selectrons for the case when the neutralino is gaugino-dominated. In
fact, the cross-sections fall as the selectron masses go up. As we
shall see presently, the predicted number of four-jet events from
neutralino pair-production is small, so it is desirable, for our
purposes, to choose the parameters to maximise the cross section.
This is achieved by choosing the selectron masses as light as
possible. We have chosen the left-selectron mass consistent with a
sneutrino mass of 60 GeV; the right selectron is also chosen to have
a mass of 60 GeV. These values are more or less at the edge of the
allowed range \cite{LEP-1.5_results}.

The decays of the LSP are mediated by right-squark exchanges and
hence one has a nominal dependence on the relevant squark masses too.
This is rather weak, however, because, the branching ratio being
unity, the only effect of increasing the squark mass is to change the
kinematic distributions of the decay products. After merging of jets,
however, much of this effect --- such as it is --- is washed out.
Consequently, the final cross-section has very little dependence on
the squark mass, especially when the value becomes significantly
larger than the neutralino mass. This is always the case if the
squark mass is taken as 150 GeV or above. In our analysis we set the 
squark mass to 300 GeV for definiteness.

At this point, it might be worth mentioning that a much lower value of
squark mass is probably consistent with CDF/D0 bounds in a scenario
in which $R$-parity is violated. Current CDF/D0 bounds 
\cite{CDF_squark_bound} on squark as well as gluino masses are derived 
from signals which trigger on missing energy and momentum and may be 
considerably relaxed \cite{DPRoy} if the LSP decays, especially in
the case where baryon-number is violated. Limits on squark and gluino 
masses from CDF/D0 data in the case of $R$-parity violation with $\l2p$
couplings have not, in fact, been investigated thoroughly, though such
an analysis is, in principle, possible \cite{DreinerGuchaitRoy}.  
In any case, for the present analysis, squark masses are set rather high,
so we have not exploited the absence of a bound.
On the other hand, we {\it have} used gluino masses in the ranges allowed
by LEP-1 without reference to CDF/D0. Of course, one could relax the 
assumption of gaugino mass unification at a high scale, in which
case the gluino mass would become irrelevant as a parameter. 

The results of our analysis of the four-jet signal arising from
neutralinos of the appropriate mass is shown in Figure 2 as a scatter
plot of the predicted number of four-jet events versus the mass of
the neutralino (LSP). Each point in the scatter plot corresponds to a
different set of $(\mg, \mu, \tb)$ in the ranges $\mg =$ 0.1 to 1 TeV
(in steps of 10 GeV), $\mu =$ -- 1 TeV to 1 TeV (in steps of 25 GeV)
for $\tb =$ 1.5, 2, 5, 10, 15, 20, 25, 30, 35 respectively, subject
to constraints arising from LEP-1 data and kinematic 
accessibility to LEP-1.5.
For this plot, we have taken the sneutrino mass to be 60 GeV for
reasons explained above and set the masses of squarks belonging to
the first two generations to $m_{\sql} = m_{\sqr} = 300$ GeV.  It may
be noted that the number (which assumes 5.7 pb$^{-1}$ integrated
luminosity at $\sqrt{s} = 130$ GeV) of four-jet events never rises
above 1.8 events which, added to the SM background of 8.6 events,
barely touches 10.4 events and never approaches anywhere near the ALEPH
observation of 16 events. It is quite clear, therefore, that the
pair-production of neutralinos followed by their $R$-parity violating
decays {\it cannot be the explanation} of the observed excess in
four-jet events. This is a fairly robust result despite the crudity
of our parton-level analysis, since it is hard to see a more refined
analysis changing the result by a factor of 4 or more, which would be 
required to explain the observed events.  \\

\begin{center}
\begin{tabular}{|c|c|c|c|}
\hline ($m_{\widetilde{g}},\mu,\tan \beta$) 
                     &(340,-400,2)   & (360,-300,15) & (370,300,30)
\\ \hline $M_{\widetilde{\chi}^0_1}$   
                     & 51.9 GeV      &  51.9 GeV    & 51.8 GeV
\\ \hline produced neutralino pairs    
                     & 6.13          &  5.75        & 5.59
\\ \hline $\Sigma \mid \!${\bf p}$\!\mid_{jet} \geq  0.1 \sqrt{s}$        
                     & 6.13          &  5.75        & 5.59 
\\ \hline $\geq 4$ jet events only            
                     & 3.32          &  3.10        & 3.01 
\\ \hline $M_{ij} \geq 25~{\rm GeV}$
                     & 2.09          &  1.96        & 1.90   
\\ \hline Multiplicity \& jet masses
                     &$\sim$1.25     & $\sim$1.18   &$\sim$1.14 
\\ \hline  
\end{tabular}  
\end{center}
\noindent
{\rm Table 1}: {\sl Illustrating the effect of various cuts on the
neutralino induced multijet signal of Figure 2 for specimen points in
the parameter space. } \\

The numbers displayed in Figure 2 may seem unexpectedly small, in view
of the fact that the neutralino production cross-section can be as
large as 1 --- 1.5 pb at LEP-1.5 \cite{neutralino_production}. The
explanation for this lies in the selection criteria imposed by the
ALEPH Collaboration.  The effects of these criteria are illustrated in
Table 1 for three specimen points in the parameter space where the
cross-section is relatively large and the neutralino mass lies around
52 GeV, which means the distribution in the sum of dijet invariant
mass would be peaked around 104 GeV. Initially, about 6
neutralino-induced events are indeed predicted. The cut removing soft
jets makes no impact on the
signal. This is easy to justify using simple kinematical arguments.
However, the channels with four or more jets retain only about half of
this cross-section. This is, in turn, further reduced by one-third by
the requirement that the dijet invariant mass of each pair of
observed jets be greater than 25 GeV. The signal is already down to 
about 2 events and can be expected to fall to barely more than 1 event by
application of cuts on charged track multiplicity and jet invariant 
mass.

Finally, it may be noted that the effects of initial state radiation
can, in general, increase the neutralino pair-production
cross-section \cite{ALEPH_4jet} because the effective centre-of-mass
energy then falls back near the $Z$-resonance; however this drives us
closer to the threshold for production of neutralinos so that the
cross-section undergoes some phase-space suppression. The net result
of these opposing effects is, in general, to keep the cross-section
just so. In any case, however, the final numbers for neutralino
production are so small that this point is merely academic.

\begin{center}
{\bf 4. FOUR-JET EXCESS FROM CHARGINO PAIR-PRODUCTION.}
\end{center}

We now turn to the other possibility that the four-jet signals arise
from the production of a pair of (lighter) charginos of mass in the
range 46 -- 65 GeV which is allowed by LEP-1 data and accessible to
LEP-1.5.  Each chargino decays to a neutralino (LSP) and an off-shell
W-boson which then goes to a pair of jets or a pair of leptons. The
neutralino then decays (as before) to three jets. Either of the
following things can happen:
\begin{enumerate}
\item Both off-shell $W^*$'s can decay to hadrons, making five jets
in all from each chargino. The ten jets in the final state can then
merge to give four-jet events.
\item One off-shell $W^*$ can decay to leptons and the other to hadrons.
The leptons can evade detection if they lie within the jets coming
from the decay of the neutralino and the other $W^*$. After merging
of jets, this configuration, too, can yield some four-jet events.
\item One off-shell $W^*$ can decay to leptons and the charged lepton
can be isolated from the jets, leading to a signal with a hard
isolated lepton (electron or muon), missing energy and multijets.
\item Both off-shell $W^*$'s can decay to leptons leading to a signal
with a pair of hard leptons (electrons or muons) of opposite sign,
large missing energy and multijets.
\end{enumerate}
Of these, the first two will contribute to the four-jet excess
observed by ALEPH. They can also, in principle, lead to multijet
events with higher multiplicity than the four- and five-jet events
studied by ALEPH, since, after all, we start with ten jets. The other
two options will lead to clear signals which should be observable not
only by ALEPH but also by the other detectors at LEP, provided the
signal is large enough. This would constitute an extra test of the
scenario under consideration, provided it proves workable in the first
place. 

Like the cross-sections for neutralino production, the cross-sections
for chargino pair-production are also well known and we have checked
that our results are in agreement, both analytically and numerically,
with those of earlier authors \cite{chargino_production}. As the
predicted cross-sections for chargino production are much larger than
those for neutralinos, we get a sizable residue after application of
all the relevant cuts.  The decays of the chargino to a neutralino
and a pair of light quarks are $R$-parity conserving and again have
been studied before \cite{Bartl_etal}. The novel feature of our
analysis is simply the decay of the neutralino (LSP) into jets.

In order to have charginos of mass in the appropriate range, we are,
as before, restricted to a part of the MSSM parameter space. This
corresponds to the region between the solid and dotted lines (for
each $\tb$) in Figure 3. We have not included CDF/D0 bounds in this
figure for reasons explained above. Thus, we consider a fairly large 
part of the parameter space which supports the relevant masses of 
the lighter chargino.

The chargino production cross-section has contributions from
$s$-channel $\gamma,Z$ exchanges and $t$-channel $\snu$ exchange.
The variation of the chargino cross-section with the mass of the
sneutrino is given, for a fixed set of other MSSM parameters, in
Figure 4.  The $s$ and $t$ channel contributions are known
\cite{chargino_production} to interfere destructively, leading to a
dip in the cross-section for some values of the sneutrino mass when
the charginos are gaugino-dominated, as is the case in this figure.
The dip arises in the region $\msnu \simeq 30 - 40$ GeV, which is
ruled out by LEP-1 data. Thus, in the allowed region, the
cross-section essentially grows with sneutrino mass, with a tendency
to saturate as the mass goes as high as a few hundred GeV. In this
work, we shall require a somewhat small chargino production
cross-section, so that it becomes desirable to choose a sneutrino
mass at the lower end. We choose 65 GeV, which is marked by a bullet
in the figure and corresponds to a cross-section of about 8 pb.

We then consider decays of the charginos to LSP's and jets and/or
non-isolated leptons, followed by decays of the LSP's to jets. This
involves, as before, a nominal dependence on the squark mass, which,
is, however, weak, as in the case of neutralino production. We set
the squark mass to 500 GeV in the subsequent analysis. The ten (or
eight) jets in the final state are passed (as in the neutralino case)
through our simple-minded jet-merging algorithm to determine the
number of distinct jets.  These results are illustrated in Figure 5
which is a scatter plot of the predicted number of four-jet events
for chargino masses in the range 45 --- 65 GeV for different values
of the MSSM parameters.  As in Figure 2, each point in the scatter
plot corresponds to a different set of $(\mg, \mu, \tb)$ in the
ranges $\mg =$ 0.1 to 1 TeV (in steps of 10 GeV), $\mu =$ -- 1 TeV to
1 TeV (in steps of 25 GeV) for $\tb =$ 1.5, 2, 5, 10, 15, 20, 25, 30,
35 subject, as before, to constraints imposed by LEP-1 data and
accessibility at LEP-1.5. Again, as beforem no CDF/D0 constraints are 
imposed. The points marked by bullets in Figure 3
are selected out of Figure 5 by imposing the conditions that the
number of four-jet events is $(16 \pm 0.5)$ and 50 GeV $<
M_{\widetilde{\chi}_1^+} <$ 55 GeV. It should be noted that the 
bullets correspond to {\bf some} value of $\tb$ among the listed
values; not necessarily one of the values marked on the contours.
Thus, points which lie in the region ruled out for $\tb >5$ correspond
to $\tb = 1.5$ or 2. 

It may be seen that the events are fairly densely clustered in an arc
which gradually goes down as the chargino mass increases. This is
indicative of phase-space suppression rather than a diminishing
coupling. For this figure, the sneutrino mass has been tuned to 65
GeV in order to ensure that the number of four-jet events consistent
with the ALEPH observation should be compatible with a chargino mass
of 50 -- 55 GeV, at least for the range where the points are most
thickly clustered. Since there are many points {\it above} this
region, it should be possible to decrease the sneutrino mass (this
decreases the cross-section as shown in Figure 4) further.  However,
if one increases the sneutrino mass, the cross-section and hence the
number of four-jet events goes up and we are then confined to just a
few points in the parameter space which would give the required mass
of the chargino and the required number of events. It may be
concluded, then, that the chargino solution to the four-jet excess
problem favours a light sneutrino with $\msnu \sim 60 - 70$ GeV
(though it does not demand this absolutely). What is important,
however, is that one {\it can} find at least a tentative explanation
for the ALEPH excess in terms of chargino pair-production.

\begin{center}
{\bf Table 2} \\[5mm]
\begin{tabular}{|c|c|c|c|}
\hline ($m_{\widetilde{g}},\mu,\tan \beta$) 
                     &(160,-600,2)    & (200,500,15)   & (200,400,30)
\\ \hline $M_{\widetilde{\chi}^\pm_1}$ 
                     & 53.7 GeV       &  54.5 GeV      &  54.3 GeV   
\\ \hline $M_{\widetilde{\chi}^0_1}$   
                     & 25.0 GeV       &  28.2 GeV      &  28.3 GeV
\\ \hline  produced chargino pairs
                     & 30.0 (9.4)     & 27.9 (9.5)     & 28.3 (9.7)
\\ \hline  $\Sigma \mid \!${\bf p}$\!\mid_{jet} \geq  0.1 \sqrt{s}$       
                     & 30.0 (9.4)     & 27.9 (9.5)     & 28.3 (9.7)
\\ \hline $\geq 4$ jet events only            
                     & 18.5 (4.8)     & 17.5 (5.2)     & 17.8 (5.4)
\\ \hline $M_{ij} \geq 25~{\rm GeV}$
                     & 12.8 (2.8)     & 12.7 (3.5)     & 12.8 (3.5)
\\ \hline Multiplicity \& jet masses
                     &$\sim$7.7 (1.7) &$\sim$7.6 (2.1) &$\sim$7.7 (2.1)
\\ \hline  + SM background
                     &$\sim$16.3      &$\sim$16.2      &$\sim$16.3
\\ \hline
\end{tabular}  
\end{center}
\noindent
{\rm Table 2}: {\sl Illustrating the effect of various cuts on the
chargino-induced multijet signal of Figure 5 for candidate points in
the parameter space. Numbers in parantheses show the contribution
from events where the final state contains a lepton which goes
undetected.} \\

In Table 2, we show the effect of different kinematic cuts for three
points in the parameter space where the cross-section is consistent
with the ALEPH observation and with a chargino mass of around 54 GeV.
These effects are rather similar to those in Table 1. We start with
about 30 events.  Once again, the cut removing soft jets makes no 
impact on the signal (the
reasons are the same as before) and the four-or-more-jet 
channels contain about 60 \% of the signal.  The
requirement that each dijet invariant mass be greater than 25 GeV
reduces this to about 43 \% and the final reduction by about 40\%
due to multiplicity and jet invariant mass cuts
brings the signal down to the required level, which is about 25 \% of
the original cross-section. Events where there is a lepton which goes
undetected because of its non-isolation from the nearest jet make up
about a quarter of the excess contribution.

It is interesting that the jet merging algorithm makes the four-jet
channel the {\it dominant} one, though some three-jet and five-jet
events are also predicted. This is illustrated in Figure 6. Just a 
single six-jet event is predicted --- which is consistent with the ALEPH
observation of none, in view of the low statistics. The
five-jet events are subsequently converted to four-jet events in our
analysis, following the ALEPH Collaboration. The three-jet events, of 
course,
have large QCD backgrounds.  The fact that the ten jets from chargino
decay merge to give multijet signals which peak for precisely
four-jets is a somewhat unexpected result and is one of our most
important observations. It is worth mentioning at this juncture that
the use of the Durham algorithm for jet-merging does not change this
conclusion.

It is also noteworthy that the region in parameter space which gives
a viable cross-section for chargino pair-production leads to a LSP
mass of 25 --- 28 GeV. As explained before, this is constrained, 
by the total $Z$-width only because of the presence of 
$\l2p$ couplings. Hence such a low value for the LSP mass will be
allowed in our scenario under currently available LEP-1 and LEP-1.5
constraints. However, it is perhaps worth mentioning that with this 
LSP mass one would predict 2.1, 7.6, 7.9, 3.6, 0.5 events in the 
1, 2, 3, 4, 5-jet channels respectively for parameters corresponding 
to the first column of Table 2 (the numbers are quite similar for the 
other two columns). These jets would not affect the signal in Table 2 
because of the lower invariant mass of the jets, but might possibly be
observable (though the QCD backgrounds would also be significant). It
would be interesting to conduct such a search within a more general
study of supersymmetry signals in the presence of $\l2p$ couplings.
Such a study has, in fact, been taken up \cite{L3andUs}.

Let us now address the important question of event shapes in the scenario
discussed in this work. The parameters are chosen such that the mass
of the produced chargino is around 52 -- 54 GeV. One should,
therefore, expect the four-jet invariant mass to peak at 105 GeV or
thereabouts. However, the peak is much smeared out because of the
jet-merging effects. There is a further smearing due to addition of
the contribution from events with a lepton and a neutrino -- in these
the neutrino carries away some energy, shifting the invariant mass
peak lower than expected from the chargino mass alone. Further smearing 
effects could come from energy rescaling and detector effects,
though these are not done here. Our final
result is illustrated in Figure 7, where the solid line represents
the four-jet events predicted in each bin of width 3.15 GeV. The
parameters are chosen to match the first column of Table 1. (We have
checked that the distribution does not change much for the other two
columns.) The distribution showed by the solid line includes the SM
distribution which is also shown separately by the dashed line. The
dotted lines show the actual data observed by ALEPH (Figure 2($a$) of
Ref. \cite{ALEPH_4jet}). It is clear that while our distribution is 
an improvement on the SM, it is too broad to be a viable 
explanation of the observed events. In fact, the distribution shown
has about 21 events, since the reduction of 60\% is not included.
With this cut, the distribution would look even flatter. The probablity
that the observed distribution is a fluctuation from our prediction
is of the order of $10^{-3}$, which is somewhat better than the SM case, 
but not large enough for this possibility to be taken seriously as an 
explanation of the ALEPH four-jet events. Accordingly we conclude that
the pair-production of charginos and their decays as conceived here
cannot explain the ALEPH four-jet anomaly. This conclusion is not as
robust as that for neutralinos, because the distributions may change
when cuts are applied on charged track multiplicity and the sum of jet
invariant masses, but it is unlikely that these will change the flat 
distribution so radically as to afford an explanation of the 
sharply-peaked data.

Finally we consider the possibility of seeing hard leptons ($e,\mu$)
and missing energy in conjunction with multijets which could be an
additional 
test of the chargino signal in the current data sample. We have
checked that barely a single event is predicted for the dilepton +
jets + missing energy signal.  Figure 8 shows the distribution of the
single lepton + jets signal in different multijet channels with a cut
of 10 GeV on the minimum energy of the lepton. The largest number is
predicted in the three-jet channel. This varies from 2.8 to 1.8 as
the isolation criterion is changed from $\Delta R$ = 0.4 to 0.6. This
channel has a larger QCD background than the four-jets channel,
which, however, has a smaller number of events. As the isolated
lepton signal is easier to detect, one may expect it to be seen in
all the detectors, so that the actual numbers should be multiplied by
a factor of about 4. It may be interesting to see if these signals
can be isolated from the SM background, which would arise principally
from $b \bar b$ production since, in most of the cases, the lepton
lies just outside the nearest jet.

In Figure 9 we exhibit the isolation of the single hard lepton from
the nearest jet for each multijet channel. As we have just noted, it 
is interesting that most of the leptons lie within this nearest jet 
or very close to it. Since most of the jets in this analysis arise from 
merging, one should expect the jets to be rather fat and hence the 
choice of the cutoff value of $\Delta R_{min}$ may be taken as 0.5 or 
even 0.6 rather than the canonical choice of 0.4. In that case it may be 
even more difficult to isolate the chargino signal from the SM background.
Any further analysis of this, however, would require a more detailed
simulation of jets and is outside the scope of the present work. 

\begin{center}
{\bf 5. SUMMARY AND CONCLUSIONS.}
\end{center}

To summarise, then, we have considered two possible explanations of
the observed excess in four-jet events in $e^+e^-$ collisions at
LEP-1.5. The pair-production of (lightest) neutralinos, followed by
their decays to three jets apiece through baryon number-violating
$\l2p$ couplings and subsequent merging of these jets to yield a
four-jet signal compatible with the observed one turns out to be a
non-starter since practically all the events are lost in the
kinematic cuts. On the other hand, the more complicated case of
(lighter) chargino pair-production and their decay to (lightest)
neutralinos and jets and/or non-isolated leptons, followed by baryon
number violating decays of these neutralinos, with jets merging as
before, yields numbers adequate to explain the observed excess.
This is essentially because the raw cross-section for chargino
pair-production in $e^+e^-$ collisions is much larger than the
corresponding cross-section for neutralinos; though a large number of
events are indeed lost through the kinematic cuts, we are left with
interesting numbers in the final analysis. However the event shapes
turn out to be too different from the observation for this option to 
be a good explanation of the four-jet anomaly.

Before concluding, we would like to point out two caveats to the results
presented in this paper. One is the obvious requirement that more
data need to be analysed so that the observed excess is put on firm
ground and we have a clearer idea about the event shape. This is more
so because the other experiments at LEP have not observed
\cite{LEP-1.5_results} any such excess. Unfortunately LEP has already
gone on to higher energy runs, so such data analyses do not seem to be
forthcoming. The other caveat is the more
technical point that this analysis requires to be repeated (for the
parameter space of interest which is mapped in Figure 4) with a more
detailed simulation of the jet kinematics and detection efficiencies
since the parton-level algorithm for jet-merging is at best
representative. 

In spite of the relative crudity of our analysis, however, we have
been able to establish that the simplest application of $R$-parity 
violation to explain the four-jet anomaly is inadequate. Though
indeed chargino pair-production can yield sufficiently large
cross-sections to give the four-jet excess, it leads to broad
distributions in the four-jet invariant mass which cannot explain
the sharply peaked distribution discovered by the ALEPH Collaboration.
A refined analysis using jet fragmentation and detector simulations
is not likely to change the qualitative result, though the actual
numbers may change. One therefore has to look for some 
other explanation of the anomaly than the one studied here. In fact, 
it seems from our analysis that any explanation which involves jet-merging 
will lead to smeared distributions and the best bet seems to be to
consider two-jet decays of each of the produced particles, whatever
they may be \cite{Debajyoti}. 
We therefore conclude on a negative note. Though
several interesting features have come up during the analysis, 
at least one candidate solution to the four-jet problem seems to be 
unacceptable.

\newpage
\begin{center}
{\bf Acknowledgements.}
\end{center}

The authors are grateful to the organisers of the Fourth Workshop on
High Energy Particle Phenomenology, Calcutta where the idea for this
work originated. They would also like to thank M. Bisset, D. Choudhury, 
M. Drees
and X. Tata for discussions and some important suggestions. DKG
acknowledges financial support from the University Grants Commission,
Government of India. The work of RMG is partially supported by a
grant (No: 3(745)/94/EMR(II)) of the Council of Scientific and
Industrial Research, Government of India, while that of SR is
partially funded by a project (DO No: SERC/SY/P-08/92) of the
Department of Science and Technology, Government of India.


\newpage
\thispagestyle{empty}

\begin{figure}[htb]
\epsffile[100 390 650 750]{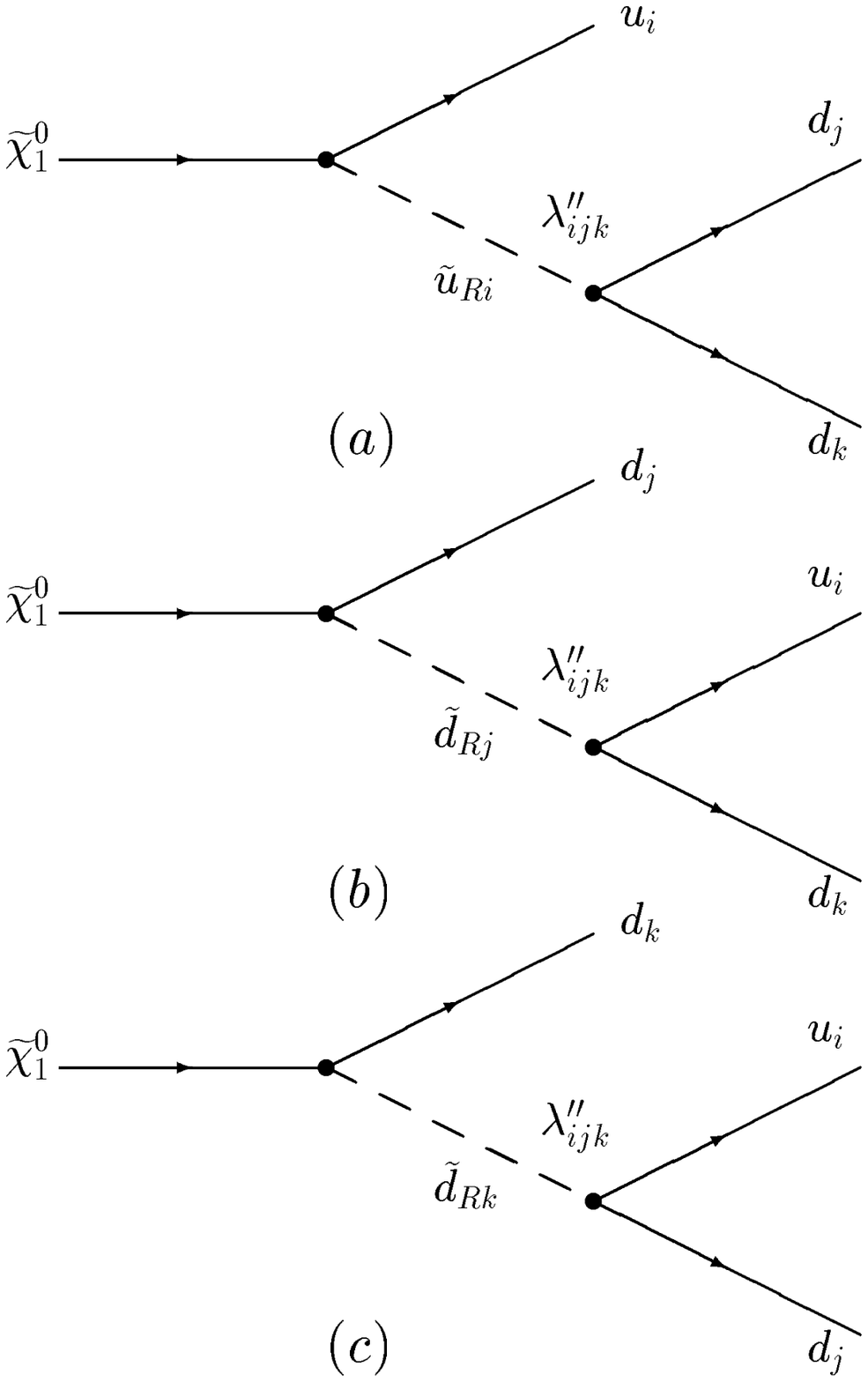}
\end{figure}
\vskip 2.5in

\noindent {\it Figure 1}. 
Feynman diagrams contributing to the decay of the LSP
($\widetilde{\chi}_1^0$) for a baryon number-violating $\l2p_{ijk}$
coupling. Only right squarks contribute to this process.

\newpage
\thispagestyle{empty}

\begin{figure}[htb]
\epsffile[100 390 432 560]{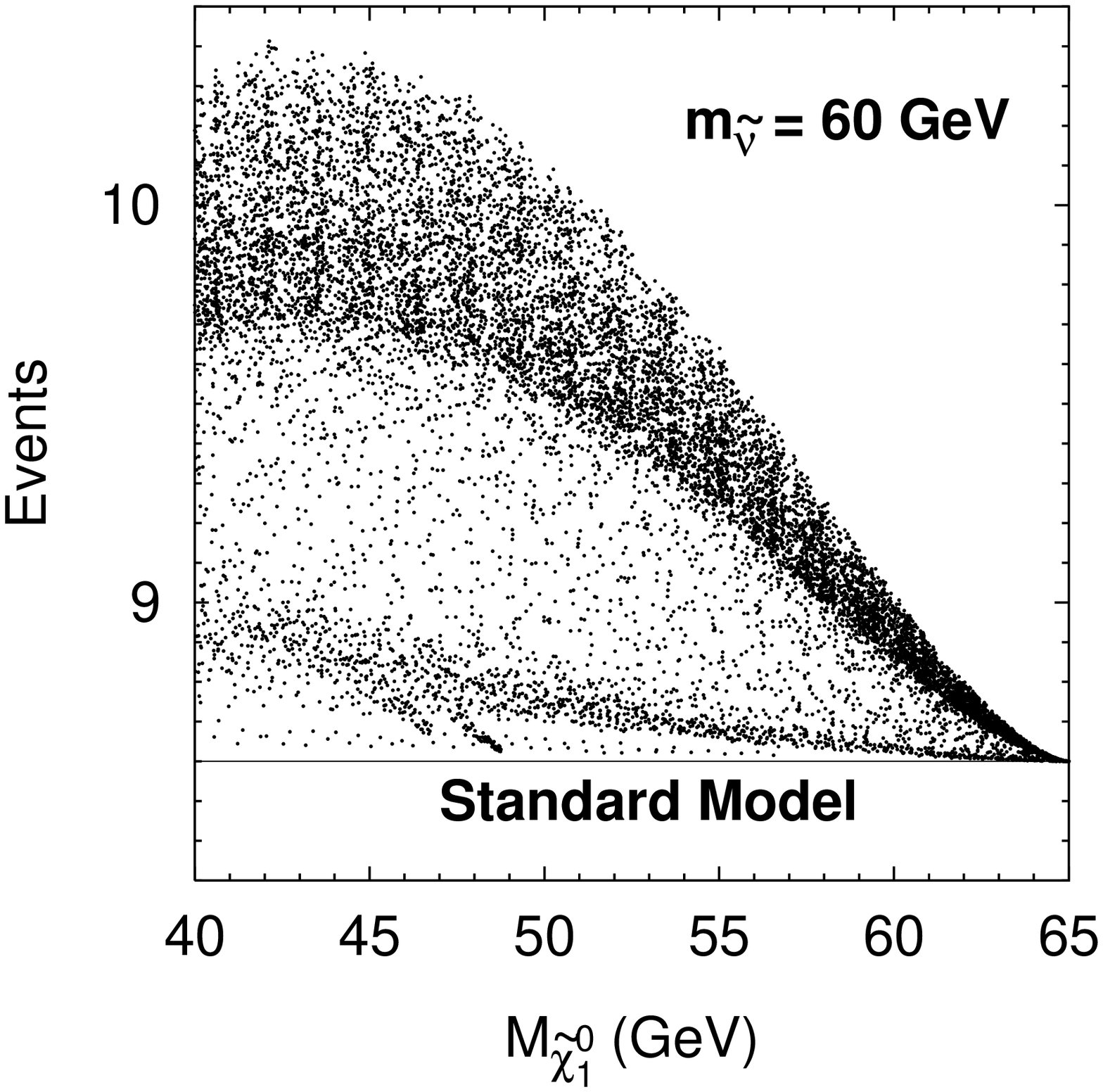}
\end{figure}
\vskip 5in

\noindent {\it Figure 2}. 
Scatter plot showing the number of four-jet events arising from
(lightest) neutralino pair-production against the mass of the
neutralino for values of $\mg,\mu,\tb$ allowed by LEP-1 constraints
(assuming 5.7 pb$^{-1}$ luminosity). We set the sneutrino and right
selectron masses to $\msnu = \mser = 60$ GeV and the squark mass
$\msq = 300$ GeV. The cross-section decreases as the former increases
and is insensitive to the latter.

\newpage
\thispagestyle{empty}

\begin{figure}[htb]
\epsffile[100 390 432 560]{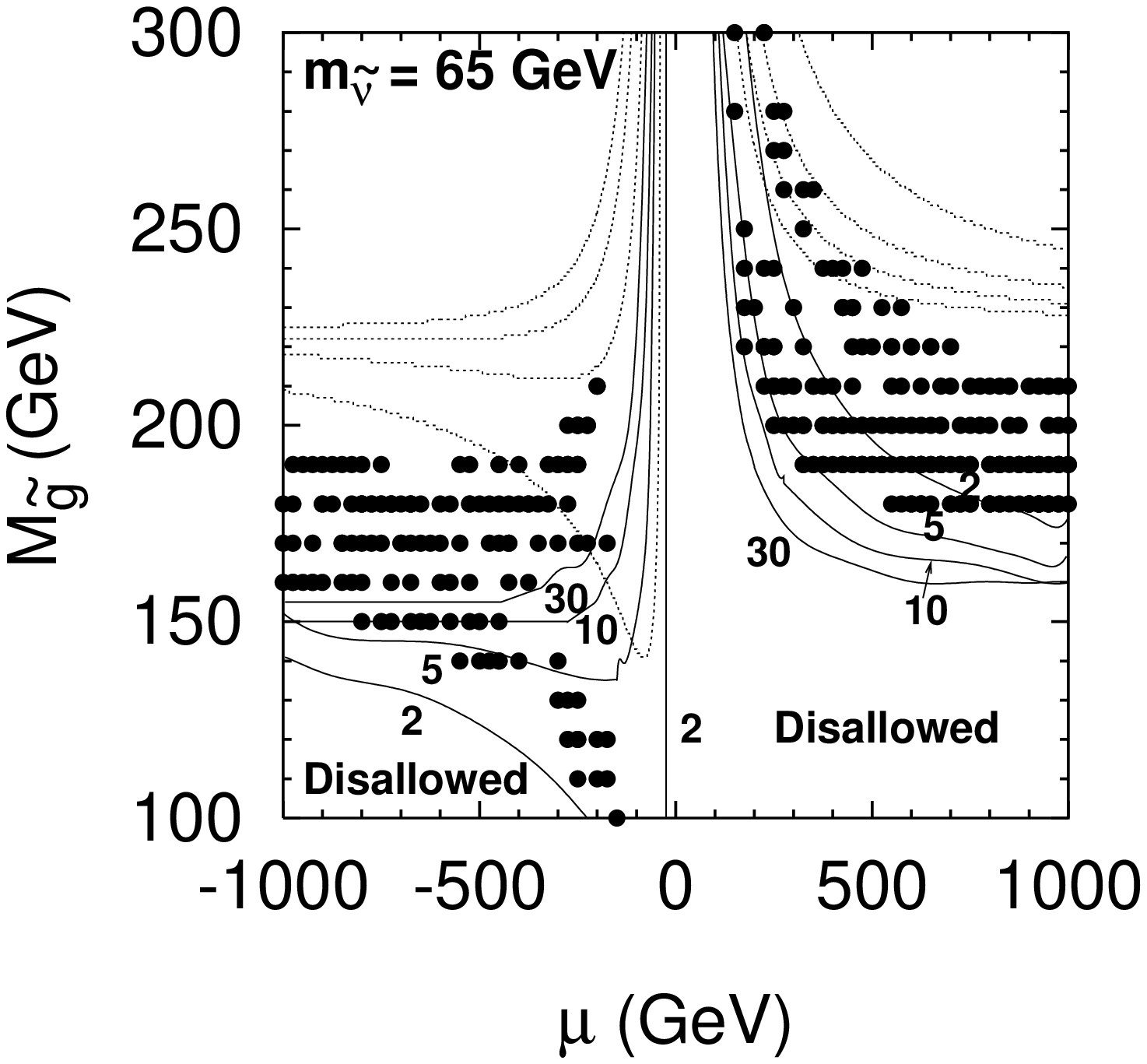}
\end{figure}
\vskip 4.5in

\noindent {\it Figure 3}. 
Bullets mark points corresponding to (16 $\pm$ 0.5) SM plus
chargino-induced four-jet events and 50 GeV $<
M_{\widetilde{\chi}^+_1} <$ 55 GeV. Each point is obtained with {\bf some}
value of $\tb$ among the set $\tb = 1.5, 2,5, 10, 15, 20, 25, 30,
35$. $\mg(\mu)$ is sampled in steps of 10 (25) GeV.  Solid lines
bound the LEP-1 excluded regions for marked values of $\tb$ and the
dotted lines bound the region kinematically accessible to LEP-1.5 for
the same values of $\tb$.  Dotted and solid curves rise (fall) with
$\tb$ in the left (right) half-plane.

\newpage
\thispagestyle{empty}

\begin{figure}[htb]
\epsffile[100 390 432 560]{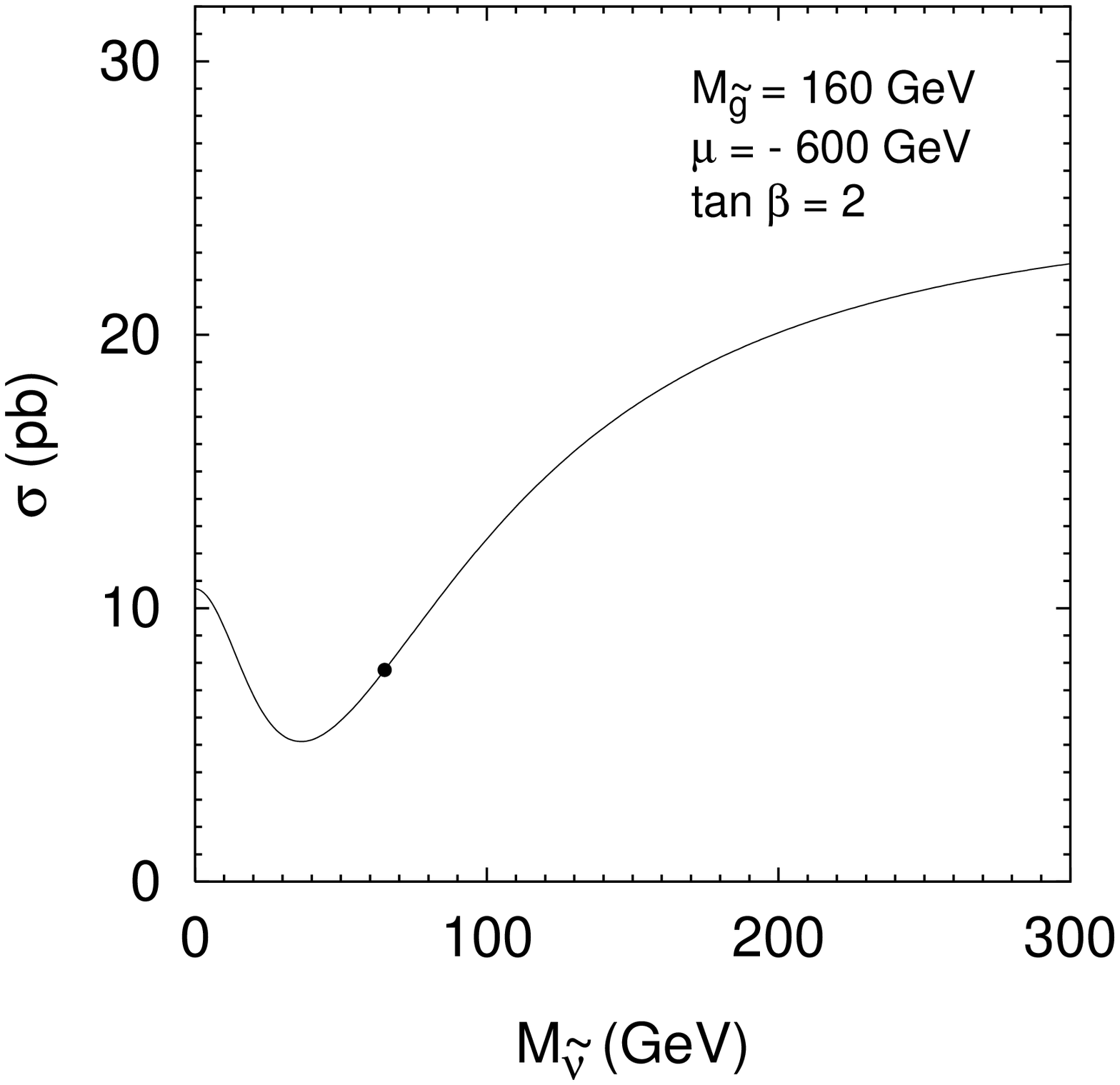}
\end{figure}
\vskip 5in

\noindent {\it Figure 4}. 
Cross-section for pair-production of the lightest chargino as a
function of sneutrino mass. The parameters are chosen to agree with
the first column of Table 2. The bullet shows the value of sneutrino
mass which yields the numbers in Figures 3 and 5 and in Table 2.

\newpage
\thispagestyle{empty}

\begin{figure}[htb]
\epsffile[100 390 432 560]{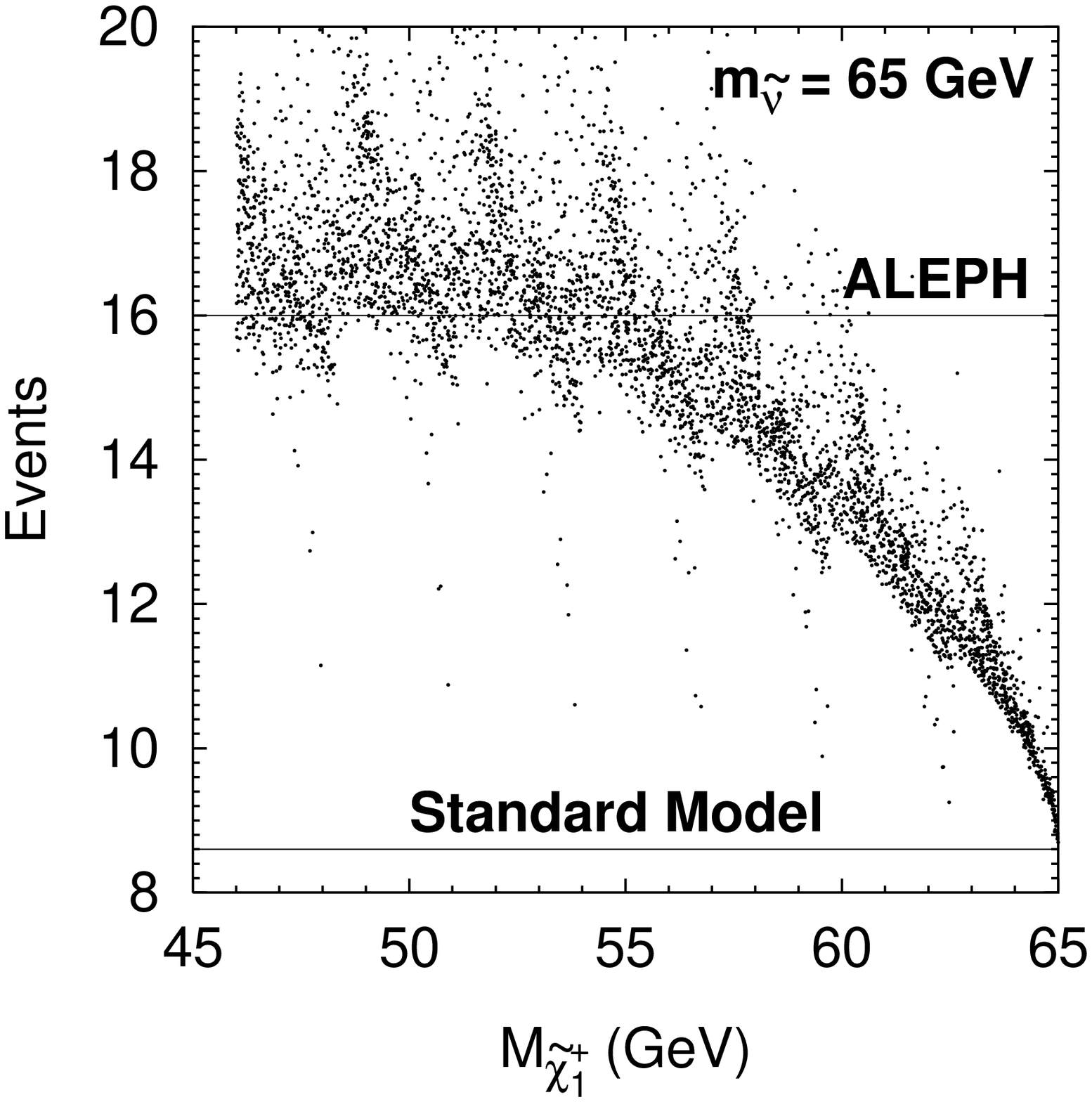}
\end{figure}
\vskip 5in

\noindent {\it Figure 5}. 
Scatter plot showing the number of four-jet events arising from
pair-production of the lighter chargino against its mass for values
of $\mg,\mu,\tb$ allowed by LEP-1 constraints. We set the sneutrino
mass to 65 GeV and the squark mass to 500 GeV.  The cross-section
increases with $\msnu$ in the allowed region and is insensitive to
$\msq$.

\newpage
\thispagestyle{empty}

\begin{figure}[htb]
\epsffile[100 390 432 560]{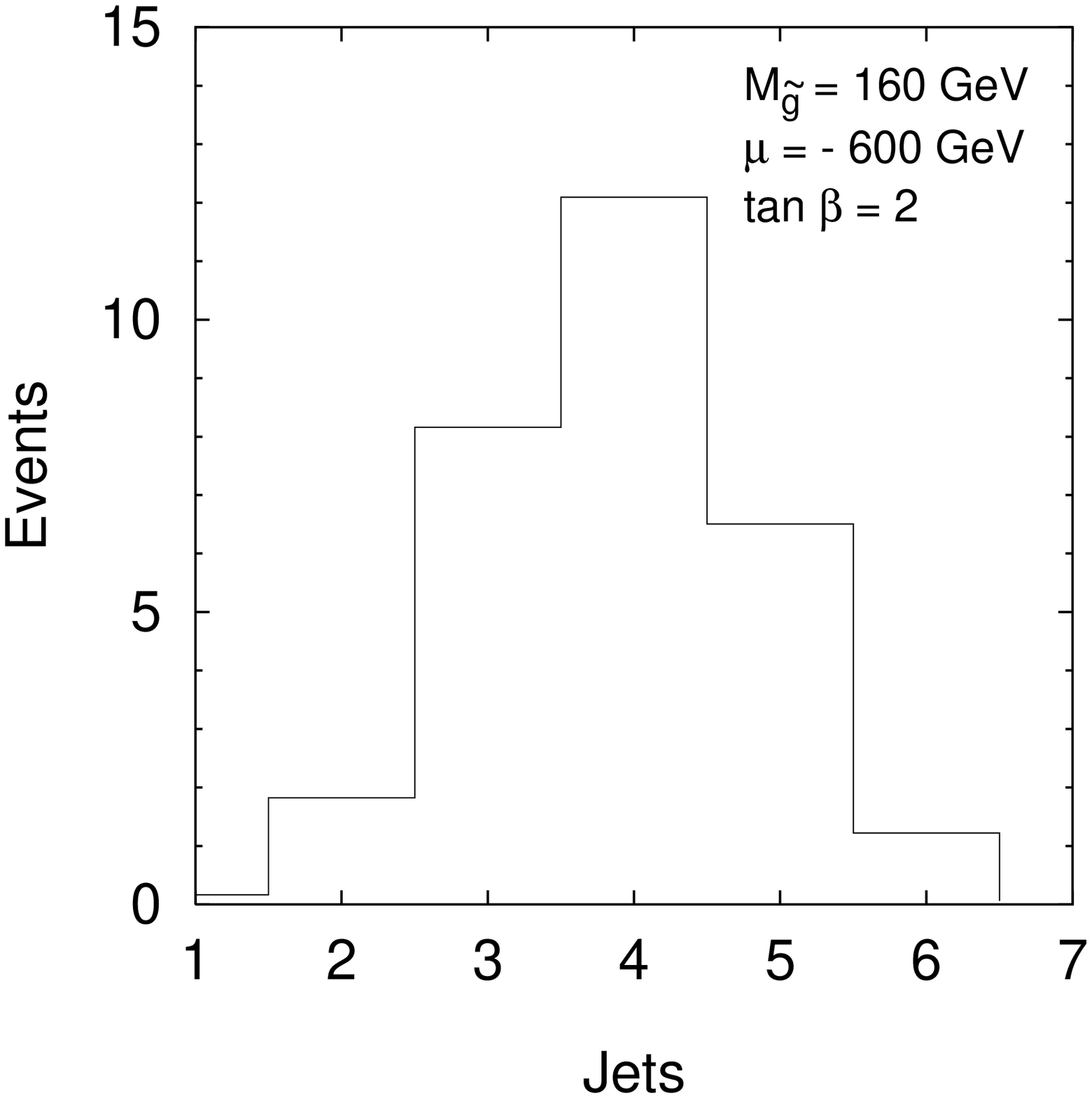}
\end{figure}
\vskip 5in

\noindent {\it Figure 6}. 
Distribution of events in multijet channels for production of
chargino pairs corresponding to the first column of Table 2 and 5.7
pb$^{-1}$ luminosity. Cuts on $M_{ij}$ and $\Delta M$ are not
imposed. The 40\% reduction assumed for charged track multiplicity
and jet invariant mass cuts is also not applied.

\newpage
\thispagestyle{empty}

\begin{figure}[htb]
\epsffile[100 390 432 560]{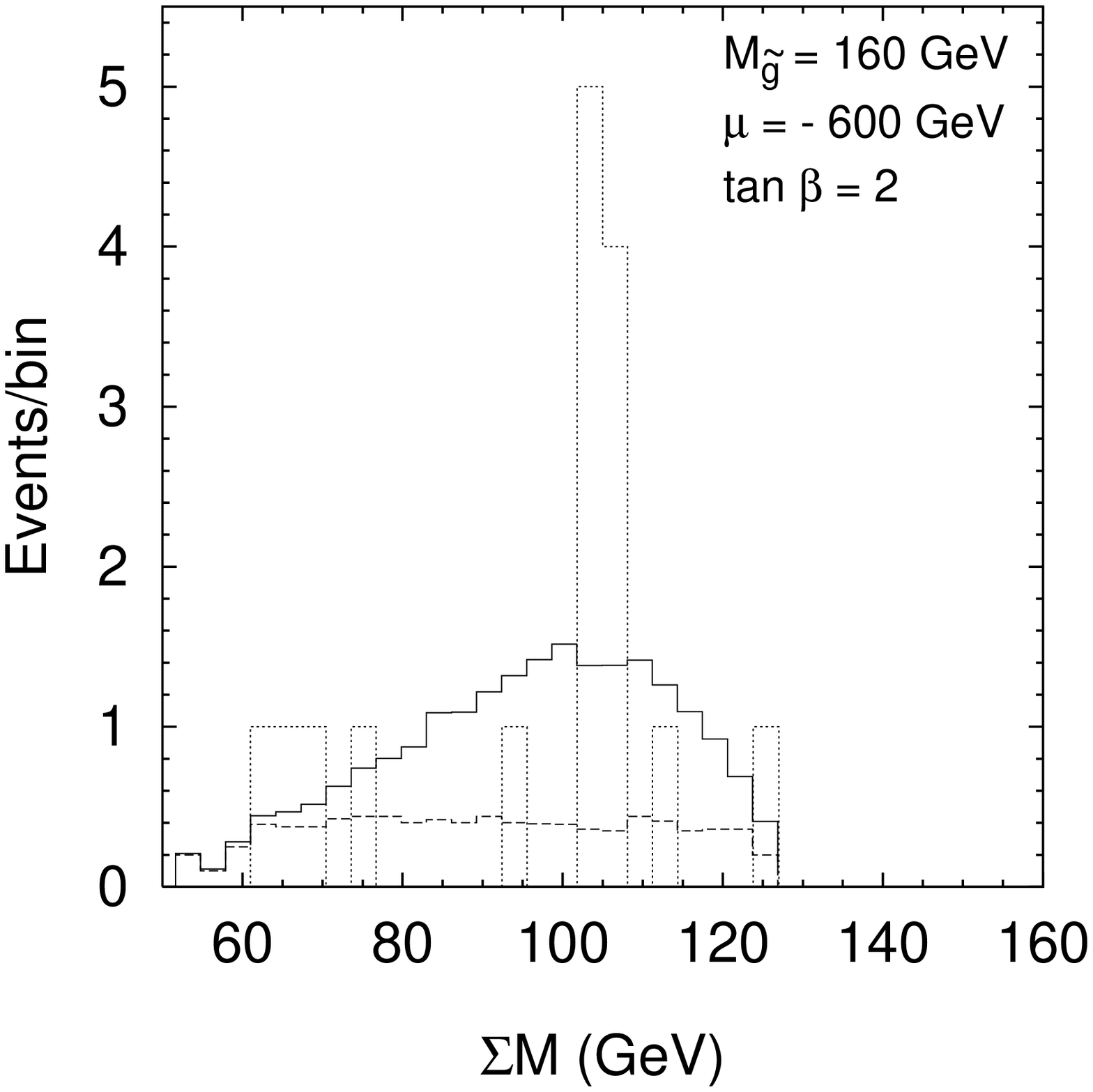}
\end{figure}
\vskip 4.5in

\noindent {\it Figure 7}. 
Distribution of four-jet events against the four-jet invariant mass
$\Sigma M$ for 5.7 pb$^{-1}$ luminosity. The solid line represents
the prediction from chargino pair-production for the parameters in
the first column of Table 2. The dotted line represents the actual
observation of the ALEPH group while the dashed line represents the
SM background (after Fig. 2 of Ref. \cite{ALEPH_4jet}).
The 40\% reduction assumed for charged track multiplicity
and jet invariant mass cuts is not applied.

\newpage
\thispagestyle{empty}

\begin{figure}[htb]
\epsffile[100 390 432 560]{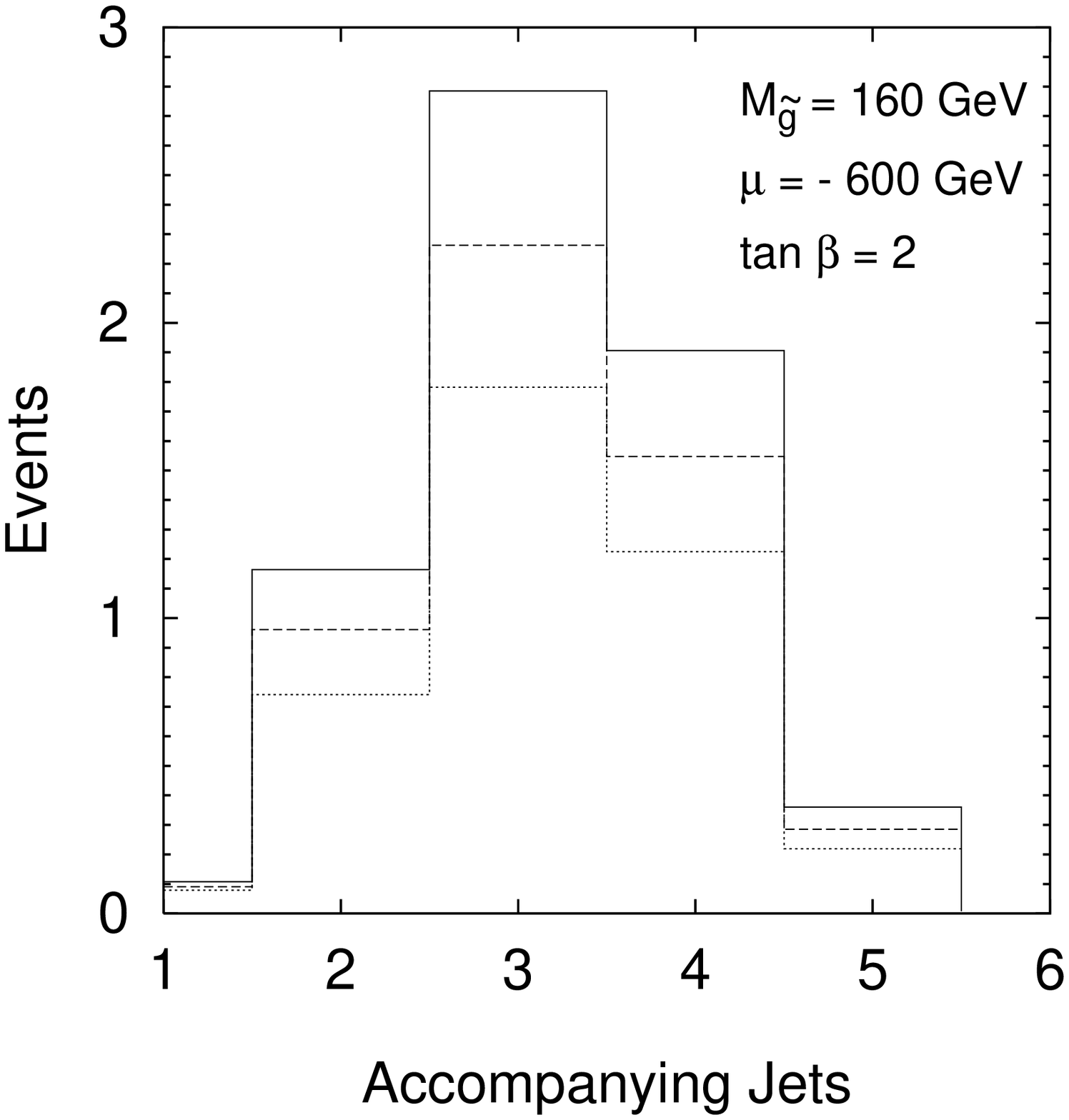}
\end{figure}
\vskip 5in

\noindent {\it Figure 8}. 
Distribution of jets accompanying a single isolated lepton for 5.7
pb$^{-1}$ luminosity for the parameters in the first column of Table
2.  Solid, dashed and dotted lines are obtained with isolation cuts
of $\Delta R >$ 0.4, 0.5 and 0.6 respectively between the lepton and
the nearest jet. There is a cut of 10 GeV on the minimum energy of
the lepton. 
\newpage
\thispagestyle{empty}
\begin{figure}[htb]
\epsffile[100 390 432 560]{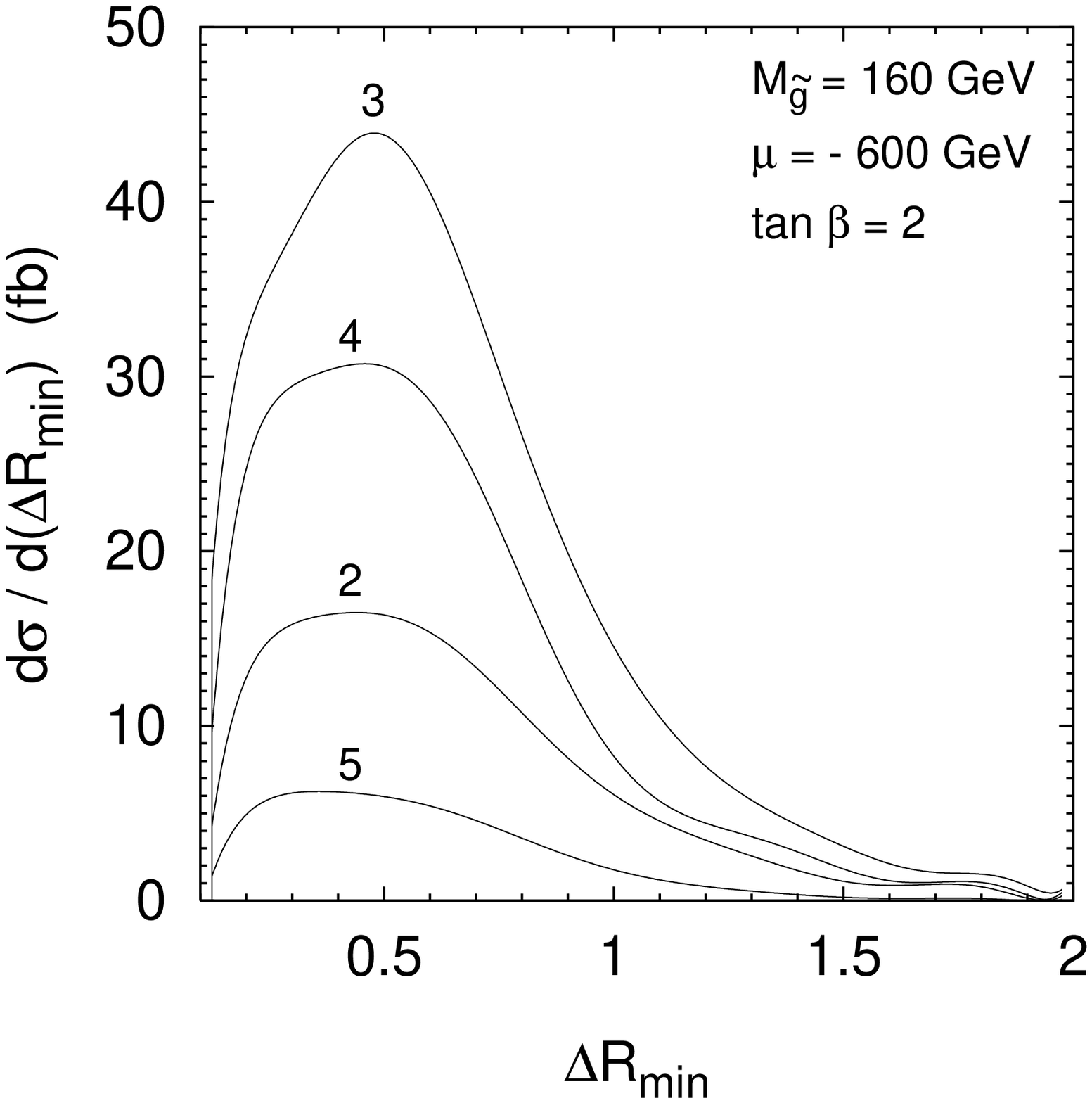}
\end{figure}
\vskip 5in
\noindent {\it Figure 9}. 
Illustrating the angular separation between a single lepton and the
nearest jet for 2,3,4,5 (marked next to the appropriate curve)
accompanying jets. Parameters are chosen as in Figure 8. There is a
cut of 10 GeV on the minimum energy of the lepton. 

\newpage
\begin{thebibliography}{99}

\bibitem{LEP-1.5_results} 
D. Buskulic \etal~(ALEPH Collaboration), \plb {\bf 373} (1996) 246 and 
CERN preprint and CERN-PPE/96-45 (1996);
P. Abreu \etal~(DELPHI Collaboration), CERN preprint CERN-PPE/96-003;
M. Acciarri \etal~(L3 Collaboration), CERN preprints CERN-PPE/95-191 and
CERN-PPE/96-29;
G. Alexander \etal~(OPAL Collaboration), CERN preprints CERN-PPE/96-020
and CERN-PPE/96-025.

\bibitem{ALEPH_4jet} D. Buskulic \etal~ (ALEPH Collaboration), CERN
preprint CERN-PPE/96-052 (1996), to be published in \zpc.

\bibitem{Rviolation_reviews} 
S. Weinberg, \prd {\bf 26} (1982) 287; N. Sakai and T. Yanagida,
\npb {\bf 197} (1982) 533.

\bibitem{Rviolation_bounds} 
J.L.Goity and M. Sher, \plb {\bf 346} (1995) 69; other bounds can be 
found in
C. S. Aulakh and R. N. Mohapatra, \plb {\bf 119} (1982) 316; 
L. J. Hall and M. Suzuki, \npb {\bf 231} (1984) 419;
S. Dawson, \npb {\bf 261} (1985) 297; 
S. Dimopoulos and L. J. Hall, \plb {\bf 207} (1987) 210; 
V. Barger, G. F. Giudice and T. Han, \prd {\bf 40} (1989) 2987; 
L. J. Hall, \mpla {\bf 5} (1990) 467;
S. Dimopoulos \etal, \prd {\bf 41} (1990) 2099;
D. P. Roy in {\it Proceedings of the Tenth DAE Symposium on High Energy
Physics}, Bombay (1992); 
G.Bhattacharyya and D. Choudhury, \mpla {\bf10} (1995) 1699;  
G.Bhattacharyya, J. Ellis and K. Sridhar, \mpla {\bf10} (1995) 1583;  
G.Bhattacharyya, D. Choudhury and K. Sridhar, \plb {\bf355} (1995) 193;  
D. Choudhury and P. Roy, TIFR preprint no.  TIFR/TH/96-12,
hep-ph/9603363 (1996);
R. M. Godbole, P. Roy, X. Tata, \npb {\bf 401} (1993) 67; 
P. D. Acton \etal~(OPAL Collaboration), \plb {\bf 313} (1993) 333; 
V. Barger, W. Y. Keung, and R. J. N. Phillips, \plb {\bf 356} (1995) 546; 
D. Buskulic \etal~(ALEPH Collaboration), \plb {\bf 349} (1995) 238;  
J. C. Rom\~ao \etal, Univ. of Valencia preprint no. FTUV/96-06, 
hep-ph/9604244 (1996); D.K.Ghosh, S. Raychaudhuri and K. Sridhar,
hep-ph/9608352.

\bibitem{LSP_decays} H. Dreiner and P. Morawitz, \npb {\bf 428} (1994) 31.

\bibitem{Dreiner} H. Dreiner, S. Lola and P. Morawitz, hep-ph/9606364
(1996).

\bibitem{proton_decay} 
See, for example, A. Yu. Smirnov and F. Vissani, ICTP preprint IC-96-16, 
hep-ph/9601387 (1996), and references therein.

\bibitem{BaerKaoTata} The stategy of considering a single $R$-parity 
violating coupling at a time was first suggested by Barger, Giudice and
Han (see \cite{LSP_decays}). The choice of $\l2p_{212}$ is suggested in
H. Baer, C. Kao and X. Tata, \prd {\bf 51}, 2180 (1995). For a recent
re-evaluation, see H. Baer, C.-H. Chen and X. Tata, hep-ph/9608221.

\bibitem{ALEPH_rviol} D. Buskulic \etal~(ALEPH Collaboration) in Ref.
\cite{LSP_decays}; G. Alexander \etal~(OPAL Collaboration) \plb {313}
(1993) 333.

\bibitem{barger_rviol} V. Barger \etal~in  Ref. \cite{LSP_decays}.

\bibitem{L3andUs} S. Banerjee {\it et al}, work in progress.

\bibitem{LEP-1_bounds} P. Antilogus \etal~(LEP Electroweak Working
Group), CERN preprint CERN-PPE/95-172 (1995).

\bibitem{BaerDreesTata} H. Baer, M. Drees and X. Tata, \prd {\bf 41},
3414 (1991).

\bibitem{cone_algorithm} G. Arnison \etal~(UA1 Collaboration), \plb 
{\bf 123} (1983) 115, {\it ibid.} {\bf 132} (1983) 214; C. Albajar
\etal~(UA1 Collaboration), \npb {\bf 309} (1988) 405; N. J. Hadley
(D0 Collaboration) D0 note 904 (Nov. 1989); 
H. J. Daum \etal, \zpc {\bf 8} (1991) 167; S. S. Snyder (D0
Collaboration), {\it Ph. D.} thesis (1995).

\bibitem{neutralino_production} 
D. A. Dicus \etal, \prl {\bf 51} (1983) 1030; 
S. Dawson, E. Eichten and C. Quigg, \prd {\bf 31} (1985) 1581; 
A. Bartl, H. Fraas and W. Majerotto, \npb {\bf 278} (1986) 1; 
X. Tata and D. A. Dicus, \prd {\bf 35} (1987) 2110; 
R. Barbieri \etal, \plb {\bf 195} (1987) 500;
M. Chen \etal, \pr {\bf 159} (1988) 201; 
H. Baer \etal, \ijmpa {\bf 4} (1989) 4111.
S. Ambrosanio and  B. Mele, \prd {\bf 53}, 2541 (1996); {\it ibid.} 
{\bf 52},3900 (1995).

\bibitem{DPRoy} D. P. Roy, \plb {\bf 283}, 270 (1992).

\bibitem{DreinerGuchaitRoy} H. Dreiner, M. Guchait and D. P. Roy, 
\prd {\bf 49}, 3270 (1994).

\bibitem{CDF_squark_bound} F. Abe \etal~(CDF Collaboration), 
\prl {\bf 76} (1996) 2006; S. Abachi \etal~(D0 Collaboration)
\prl {\bf 75}, 618 (1995).

\bibitem{chargino_production} 
H. Baer \etal, in Ref. \cite{neutralino_production}; 
A. Bartl, H. Fraas and W. Majerotto, \zpc {\bf 30} (1986) 441; 
M. Chen \etal, in Ref. \cite{neutralino_production}; 
D. A. Dicus \etal, in Ref. \cite{neutralino_production};
S. Dawson \etal, in Ref. \cite{neutralino_production}; 
P. Nath, R. Arnowitt, and A. Chamseddine, Harvard Report HUTP
--83/A077 (1983);
V. Barger \etal, \plb {\bf 131} (1983) 372; 
J. Ellis \etal, \plb {\bf 127} (1983) 233, {\it ibid.}, {\bf B132}
(1983) 463.

\bibitem{Bartl_etal} See Bartl \etal, in Ref.
\cite{chargino_production}.

\bibitem{Debajyoti} P. H. Chankowski, D. Choudhury and S. Pokorski, 
hep-ph/9606415; D. Choudhury and D.P. Roy, CERN preprint CERN-TH-96-203, 
hep-ph/9608264.

\end{thebibliography}
\end{document}